\documentclass[11pt]{article}
\usepackage{epsfig,subfigure}

\begin{document}

\begin{center}
{ \Large\sc High Impact but Less Popular}
\vskip 0.7cm

{T. Mart}

{\it Departemen Fisika, FMIPA, Universitas Indonesia, Depok 16424, Indonesia}
\end{center}

\vskip 0.7cm

\begin{abstract}
By using the SLAC-SPIRES database it is shown 
that, in spite of obtaining a high impact, Nature and Science are still
less popular in the particle physics community.
\end{abstract}

\vskip 0.7cm

In 2004 the Institute for Science Information (ISI) released a list of ``ten hottest
journals'' in physics \cite{phys_watch}. Those journals were ranked according to
the number of averaged citations per paper they have obtained from 
1992 until 2002. For the sake of the present discussion, we only display five of them 
in Table\,\ref{isi_5}, from which it is obvious that Science and Nature are crowned 
as the two top journals having the highest citation numbers per paper and, 
consequently, the highest impact factors 
(IF). In fact, the number of citations per paper obtained by Science in this period 
is almost seven times higher than that obtained by Physical Review D, 
one of the longstanding respected journals in particle physics. 

\begin{table}[!b]
  \begin{center}
    \caption{\small Five out of ten hottest journals in the physics subjects between 1992-2002 according
      to the ISI analysis.  Data are taken from \cite{phys_watch}.}
    \renewcommand{\arraystretch}{1.2}
    \label{isi_5}
{\small
    \begin{tabular}{clrcr}
      \hline\hline
      ~~Rank~~&~~~~Journal   & Total~ & ~~~~~~Averaged~~~~~~ & ~~IF~~~~ \\[-0.5ex]
      & & papers & citations &(2003)~ \\
      \hline
      1 & Science & 839 & 77 &29.162~~\\
      2 & Nature & 943 & 70 &30.979~~\\
      5 & Nuclear Physics B &10,740 & 14 &5.297~~\\
      6 & Physics Letters B&16,751& 13 &4.066~~\\
      8 & Physical Review D~~~~~~~~~~~~~~~~~~~~~~& 18,074&~~~~12~~~~ &~~~~4.599~~\\
      \hline\hline
    \end{tabular}}
  \end{center}
\end{table}

IF has become a source of controversial debates lately. Controversial because 
the IF definition, namely the number of citations obtained in a specific year 
by papers published in a journal in the previous two years divided by the 
number of these papers, is found to have some flaws \cite{tmart_symmetry}. 
However, this factor
would have never been seriously considered except until some people started to 
use it to control 
scientific enterprises, such as assessing the quality of individual papers, 
scrutinizing an applicant's track-record when considering research funds  
or academic promotions, and evaluating the performance of a research 
institute or a department.

As a natural consequence, some researchers feel the pressure to publish 
their papers in a journal with a high IF. Indeed, there has been a report 
that publishing papers in journals with IF above 5 is a prescription to 
get a tenure in certain universities in the US, whereas to get a PhD at 
some universities in China a graduate student must publish at least two 
papers in the journals with an IF of 4 or more \cite{monastersky}. Furthermore,
some scientists believe that publishing a paper in a high IF
journal will guarantee a high number of citation (the so-called 
``free ride hypothesis'').

In view of this, it is very important to raise such a question: does IF
(or averaged citations) really mean everything about the journal or paper
quality? To answer this question let us look back to Table\,\ref{isi_5}.
Comparing the two top journals (Science and Nature) with the rest three, 
which are considered as the habitat of the particle physics community, 
reveals the fact that both Science and Nature suffer from the 
problem of low productivity. 
Presumably, this is originated from the very strict editorial and refereeing 
processes they have used. Such mechanism might raise some complains from 
prominent scientists who sometimes
have speculative papers. It is widely known that in many branches of science
most important findings, which eventually led to Nobel prizes, came from a 
revolutionized idea which did not follow the mainstream. Such an idea would 
be most likely rejected by the current system of refereeing process, unless 
the editors and referees slightly relaxed their criteria. Take for example
the history of quantum mechanics or the finding of anomalous magnetic moment 
by Otto Stern \cite{drechsel}. Nevertheless, in spite of the fact that 
the low productivity has a negative impact on the popularity of the journal,
the IF definition (namely the averaged citations) clearly penalizes high productivity. 
This was pointed out by Jorge E. Hirsch and also became one of the 
reasons why he proposed the ``h-index'' as an alternative
measure of the scientific output of a researcher \cite{hirsch}.

\begin{figure}[!]
\centering
\begin{tabular}{c@{\qquad}c}
 \mbox{\epsfig{file=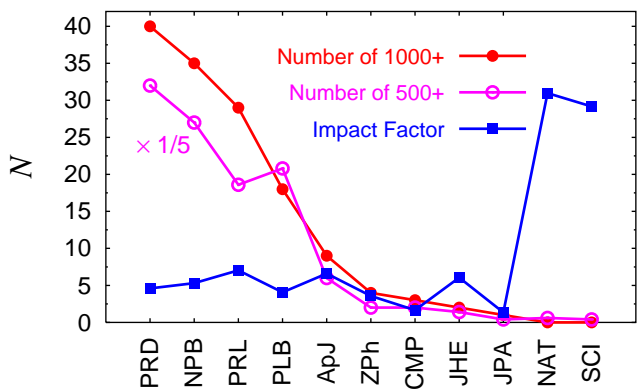,height=5.8cm}}
  \end{tabular}
\caption{\small (Color) No obvious corelation. Comparison between the number of papers with 
  more than 1000, 500 citations and the 2003 impact factor as a function of journals.
  Not all journals listed in Table\,\ref{if_1000} are shown in this figure. Review
  journals are omitted.
  \label{isi_if_compare}}
\end{figure}

To be more quantitative let us consider the particle physics papers 
with more than 1000 citations (from now on will be labeled by 1000+) 
detected by SLAC-SPIRES \cite{slac}, a database consisting of more than 
500.000 nuclear and particle physics papers published between 1950 and 2005. 
Although the database only includes Nobel papers in that period, a 1000+ paper 
obviously reflects a revolutionized idea or contains 
spectacular experimental data. The corresponding journals which published those
papers are shown in Table\,\ref{if_1000}. It is quite surprising that none
of the 1000+ papers was published in Nature and Science. Only if we go down
to 500+ papers we find that 3 papers were published in Nature and 2 papers were
published in Science. This becomes an obvious indication that the number 
of citations per paper which determines the number of IF is not 
directly correlated to the importance of journals, in the sense of 
journals that publish important papers. To visualize this fact we plot 
the number of 1000+ and 500+ papers as a function of different 
journals in Fig.\,\ref{isi_if_compare}, where as a comparison
we also show the 2003 IF of the corresponding journals.\footnote{The choice of
the 2003 IF is trivial. However, since the IF does not dramatically fluctuate
from year to year, for the purpose of the current discussion this is considered to be 
sufficient. Review journals (such as Review of Modern Physics) tend to have a
high IF.} From this figure we can directly conclude that there is no 
obvious correlation between high IF and high number of citations. 

\begin{table}[htbp]
\renewcommand{\arraystretch}{0.9}
\begin{center}
\caption{\small Number of papers with more than 1000 and 500 citations published in different
journals \cite{slac} and the corresponding IF. 
The listed abbreviations are used in Fig.\,\ref{isi_if_compare}.}
{\small
\begin{tabular}{clccr}
\hline\hline\\[-2ex]
No&Journal (Abbreviation)   & Total& Total &  IF~~~ \\
& &  ~~~1000+~~~ &~~~500+~~~ & ~~(2003) \\
\hline\\[-2ex]
1 & Physical Review D (PRD)& 40 &160& 4.599\\
2 & Nuclear Physics B (NPB)& 35 &135& 5.297\\
3 & Physical Review Letters (PRL)& 29 &93& 7.035 \\
4 & Physics Letters B (PLB)& 18 &104& 4.066\\
5 & Astrophysics Journal (ApJ)& 9 &30& 6.604\\
6 & Zeitschrift f\"ur Physics (ZPh)& 4 &10& 3.580\\
7 & Computer Physics Communication & 4 &7& 1.170\\
8 & Review of Modern Physics & 3 &16& 28.172\\
9 & Communication in Mathematical Physics (CMP)& 3 &10& 1.650\\
10 & Soviet Journal of Nuclear Physics & 3 &6& -~~~ \\
11 & Advances in Theoretical and Mathematical Physics & 2 &3& -~~~ \\
12 & Annals of Physics & 2 &11& 2.525\\
13 & Journal of High Energy Physics (JHE)& 2 &7& 6.057\\
14 & Astronomical Journal & 2 &6& 5.647\\
15 & Astrophysics Journal Supplement & 2 &5& 6.247\\
16 & Progress of Theoretical Physics & 2 &5& 2.188\\
17 & Nuovo Cimento & 2 &4& 0.285\\
18 & Proceedings of the Royal Society London & 2 &3& 1.210\\
19 & Soviet Physics JETP & 2 &2& 1.156\\
20 & Journal of Physics A (JPA)& 1 &2& 1.357\\
21 & Physica A & 1 &1& 1.180\\
22 & Nature (NAT)& 0 & 3 & 30.979\\
23 & Science (SCI)& 0 & 2 &29.162\\
\hline\hline
\label{if_1000}
\end{tabular}}
\end{center}
\end{table}

\begin{figure*}[!]
\centering
\begin{tabular}{c@{\qquad}c}
 \mbox{\epsfig{file=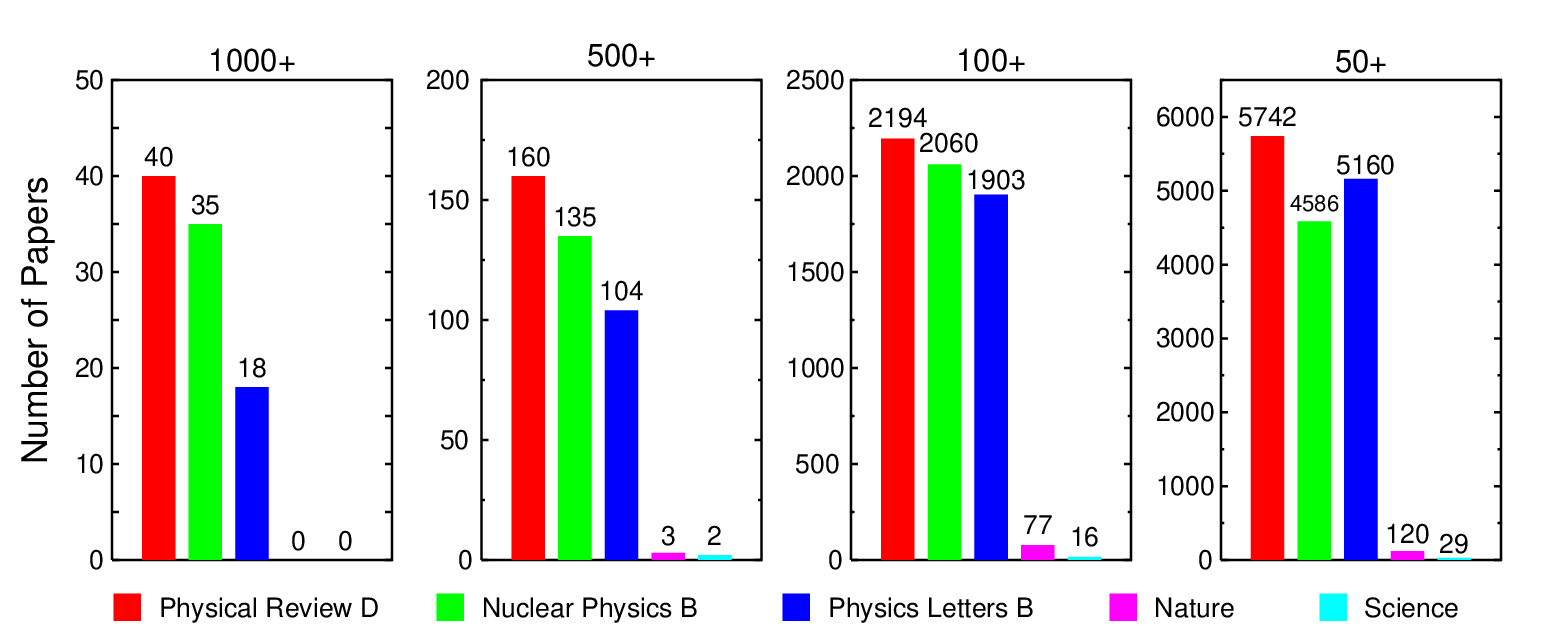,height=6cm}}
  \end{tabular}
\caption{\small (Color) The number of papers with more than 1000, 500, 100, and 50 citations 
  obtained by Physical Review D, Nature, and Science. Data are taken from \cite{slac}.
  \label{nature_prd}}
\end{figure*} 

To further investigate why both Nature and Science obtained the highest averaged
citations, in Fig.\,\ref{nature_prd} we plot the number of 1000+, 500+, 100+, 
and 50+ papers published in all five journals listed in Table\,\ref{isi_5}.
We again find a similar pattern, namely very small fractions of those highly cited
papers were published in Nature and Science. This clearly shows that, on
average, all papers published in both journals (note that they publish all physics
subjects) have citations between 50 and 100, thus yielding an averaged citation 
around 70.

As discussed above a very strict editorial process will most likely filter out 
very important publications. Besides that, we will not exclude the contribution from
the intrinsic problem of a general-reader journal. It is widely known that most
important findings usually require a very detailed technical explanation and, as a consequence,
the corresponding reports will certainly not suitable for publication in 
Nature\footnote{In 2005 the Nature Publishing Group released a new journal called
Nature Physics.}
or Science. Another obvious example is shown by Table\,\ref{if_1000}. Although
highly respected as a top American Physical Society journal, Physical Review Letters
occupies the third rank after Physical Review D and Nuclear Physics B.

To conclude, the above discussion corroborates the finding of  Per O. Seglen 
that the ``free ride hypothesis'' is really a myth \cite{po_seglen}. 
By observing two groups of scientific authors who published their papers 
in two different journals with significantly different IF, he arrived at a 
conclusion that article citation rates determine the journal IF, but not vice versa.

\end{document}